# Viral Privacy: Contextual Integrity as a Lens to Understand Content Creators' Privacy Perceptions and Needs After Sudden Attention

Joseph S. Schafer*

Human-Centered Design & Engineering, University of Washington; Center for an Informed Public, University of Washington, schaferj@uw.edu

Annie Denton

Paul G. Allen School of Computer Science & Engineering, University of Washington; Center for an Informed Public, University of Washington, adento@uw.edu

Chloe Seelhoff

University of Washington, cjseel3@uw.edu

Jordyn Vo

University of Washington, jvo2@uw.edu

Kate Starbird

Human-Centered Design & Engineering, University of Washington; Center for an Informed Public, University of Washington, kstarbi@uw.edu

When designing multi-stakeholder privacy systems, it is important to consider how different groups of social media users have different goals and requirements for privacy. Additionally, we must acknowledge that it is important to keep in mind that even a single creator's needs can change as their online visibility and presence shifts, and that robust multi-stakeholder privacy systems should account for these shifts. Using the framework of contextual integrity, we explain a theoretical basis for how to evaluate the potential changing privacy needs of users as their profiles undergo a sudden rise in online attention, and ongoing projects to understand these potential shifts in perspectives.

---

* Corresponding author

## 1 INTRODUCTION

*"[T]urning my personal Twitter into a rehashed joke as a form of protest was not something I envisioned taking off like this. If I had, I'm not sure I'd have done it at all. Prior to this, I had around 400 followers steady, a minor audience with which I was able to get some of my work out there and tweet controversial Doctor Who hot takes. As I write this, I am at 77.4K followers, all waiting with bated breath (I assume) for my next Italian [Elon] Musk tweet."* - Audrey Armstrong, from *"Mamma Mia: On Accidentally Going Viral"* [1], content in brackets added by authors.

When we consider designing for multi-stakeholder privacy, this workshop's framing rightfully acknowledges that conceptualizations of 'users' as unified and monolithic are simplistic and fail to capture important needs for various user groups. However, only acknowledging variations between user groups still presumes that a particular user's privacy needs are relatively static and cohesive. In this workshop paper, we argue that this assumption is not always the case, particularly as users' audiences on online social media platforms change. We outline the theory of contextual integrity as a possible explanatory model for these changing needs, and outline two ongoing projects on understanding the experiences of social media users who receive sudden social media attention.

## 2 CONTEXTUAL INTEGRITY

Contextual integrity was a theory in privacy law originally developed by Helen Nissenbaum in 2004, to explain under what circumstances we are likely to feel that our privacy is violated, and as a new mechanism through which to criticize widespread public surveillance [9]. This theory argues that all interactions consist of information flows, and that privacy violations occur when norms for sharing information within a particular context are violated. In Nathan Malkin's summary article of contextual integrity, he describes these contexts as having five aspects, which correspond to what kind of information is being shared, whom the data is about, who is sending the data, who is receiving the data, and how the data is being sent [7].

Prior research into social media has used the lens of contextual integrity to evaluate the privacy concerns of designs. These include Pan Shi et. al.'s analysis of Facebook friendship pages [10], Natalia Criado and Jose M. Such's design of a tool to enable implicit contextual integrity-protecting privacy tools [2], Kenneth C. Werbin et. al.'s analysis of Facebook privacy violations for users with fluid, intersectional identities [13], and Such and Criado's exploration of applying contextual integrity to multiparty information sharing [12]. Research into other related phenomena, such as context collapse, has also gained prominence, e.g. [5,8]. Furthermore, Malkin explicitly calls for more engagement with contextual integrity in both research and in practice, to better understand and protect privacy. Research has also begun to study what happens to users who gain sudden social media attention, e.g. [3,4], though these have primarily drawn from research traditions like conditioning through feedback mechanisms [11] or social impact theory [6], rather than contextual integrity.

## 3 ONGOING RESEARCH

To address this gap in research into the experience of users who experience sudden social media attention, we outline two ongoing projects that the authors in this paper are conducting.

First, to understand how creators experience events of sudden social media attention, we are recruiting and interviewing adult US-based English-language TikTok users, including how they consider content creation and self-presentation differently now that they have experienced a sudden burst of engagement. We have so far completed five interviews of TikTok creators who received a sudden influx of social media attention on the platform. Though most have not directly mentioned privacy concerns, participants have described a variety of ways in which they either experienced invasive-feeling behaviors from audiences or how they acted to preserve their privacy as the video went viral. These behaviors have included deleting past content or creating new accounts for videos they anticipated could go viral. Some participants have also acknowledged other concerns relevant to privacy as part of suddenly gaining social media attention, such as audience members doing invasive analyses of their video backgrounds and past videos, feeling uneasy about the creation of fan accounts, or privacy concerns raised by other people (such as friends or family members) appearing in their content. It is important to note that while these users have all undergone shifting contexts, their approach to navigating these increases in attention have not always embodied reducing their online presence or making themselves more private. Some participants have expressed feeling less concerned about curating their content after receiving increased attention, for example, or even described deliberately posting controversial material so that their engagement increases. As we recruit and conduct more interviews with this population, we hope to add further nuance to how privacy concerns manifest as content creation contexts rapidly shift for users receiving sudden social media attention.

Alongside our TikTok interview study, Joseph S. Schafer and Kate Starbird have been conducting a parallel research project on how people change their posting and self-presentation behaviors on Twitter after gaining a sudden increase in attention, similar to the work of [3,4]. Like the users who gained a sudden increase in attention on TikTok, these Twitter users had dramatically higher engagement with their content at this time than others, which resulted in their context of platform content creation shifting significantly. While preliminary results indicate that users did not significantly change their posting frequency long-term, users in this new context did much more frequently update their self-presentation on the platform, indicating that they are presenting themselves differently in this newer context.

## 4 CONCLUSION

When designing for multi-stakeholder privacy, it is important to note not only how different users have different privacy needs, but as contexts change for a particular user their privacy needs and concerns may also shift. One such shift that may be critical for privacy needs but remains underexplored is for users who experience sudden increases in social media attention. The framework of contextual integrity can be useful in thinking through the privacy needs of this user population, and how to design systems that adapt for individual users' changing needs.

**ACKNOWLEDGEMENTS**

We are grateful for support from the National Science Foundation, from grant DGE-2140004 and grant #1749815. We are also supported in part by the University of Washington Center for an Informed Public, the John S. and James L. Knight Foundation, and the William and Flora Hewlett Foundation. Any opinions, findings, and conclusions or recommendations expressed in this material are those of the authors and do not necessarily reflect the views of the above supporting organizations or the National Science Foundation.